\documentclass[longauth]{aa} 
\usepackage{graphicx}
\usepackage{txfonts}
\usepackage[]{natbib}
\usepackage[]{subfig}
\usepackage{color}
\usepackage{amsmath}
\usepackage{hyperref}
\usepackage{ulem}
\definecolor{gris}{gray}{0.5}

\begin{document} 

   \title{ALMA observations of the \object{Th\,28} protostellar disk}

   \subtitle{A new example of counter-rotation between disk and optical jet}

   \author{F. Louvet\inst{1}
       \and 
           C. Dougados\inst{2,1,3}
       \and 
           S. Cabrit\inst{4,3}
       \and 
           A. Hales\inst{5,6}
       \and 
           C. Pinte\inst{2,1,3}
       \and 
           F. M\'enard\inst{2,1,3}
       \and 
            F. Bacciotti\inst{7}
       \and 
          D. Coffey\inst{8}
       \and
           D. Mardones\inst{1}
       \and 
          L. Bronfman\inst{1}
       \and
        F. Gueth\inst{9}
      }

   \institute{Departamento de Astronomia de Chile, Universidad de Chile, Santiago, Chile
              \email{flouvet@das.uchile.cl}
          \and
             UMI-FCA, CNRS/INSU, France (UMI 3386)            
         \and
              Univ. Grenoble Alpes, CNRS, IPAG, F-38000 Grenoble, France 
        \and
             Laboratoire d’Etudes du Rayonnement et de la Matière en Astrophysique et Atmosphères (LERMA) - Observatoire de Paris-Meudon, France            
          \and
          Atacama Large Millimeter/Submillimeter Array, Joint ALMA Observatory, Alonso de C\'ordova 3107, Vitacura 763-0355, Santiago - Chile 
          \and
            National Radio Astronomy Observatory, 520 Edgemont Road, Charlottesville, Virginia, 22903-2475, United States
          \and 
             Instituto Nazionale di Astrofisica-Osservatorio Astrofisico di Arcetri, Firenze, Italy
          \and
          School of Physics, University College Dublin, Belfield, Ireland
          \and
          Institut de Radioastronomie Millimétrique-Grenoble
           }

   \date{Received February 2016; accepted March 2016}

 
  \abstract
   {}
   {Recently, differences in Doppler shifts across the base of four close classical T Tauri star jets have been detected with the HST in optical and near-ultraviolet (NUV) emission lines, and these Doppler shifts were interpreted as rotation signatures under the assumption of steady state flow. To support this interpretation, it is necessary that the underlying disks rotate in the same sense. Agreement between disk rotation and jet rotation determined from optical lines has been verified in two cases and rejected in one case. Meanwhile, the near-ultraviolet lines, which may trace faster and more collimated inner spines of the jet than optical lines, either agree or show no clear indication. We propose to perform this test on the fourth system, \object{Th\,28}.}
   {We present ALMA high angular resolution Band 7 continuum, $^{12}$CO(3-2) and $^{13}$CO(2-1) observations of the circumstellar disk around the T Tauri star \object{Th\,28}.}
   {The sub-arcsecond angular resolution (0.46\arcsec$\times$0.37\arcsec) and high sensitivity reached  enable us to detect, in CO and continuum, clear signatures of a disk in Keplerian rotation around \object{Th\,28}. The $^{12}$CO emission is clearly resolved, allowing us to derive estimates of disk position angle and inclination. The large velocity separation of the peaks in $^{12}$CO, combined with the resolved extent of the emission, indicate a central stellar mass in the range 1-2\,M$_\odot$. The rotation sense of the disk is well detected in both $^{13}$CO and $^{12}$CO emission lines, and this direction is opposite to that implied by the transverse Doppler shifts measured in the optical lines of the jet.}
   {The \object{Th\,28} system is now the second system, among the four investigated so far, where counter-rotation between the disk and the optical jet is detected. These findings imply either that optical transverse velocity gradients detected with HST do not trace jet rotation or that modeling the flow with the steady assumption is not valid. In both cases jet rotation studies that rely solely on optical lines are not suitable to derive the launching radius of the jet.}


   \maketitle

\section{Introduction}
\label{s:intro}

\begin{figure*}[htb!]
\subfloat{\includegraphics[scale=0.32]{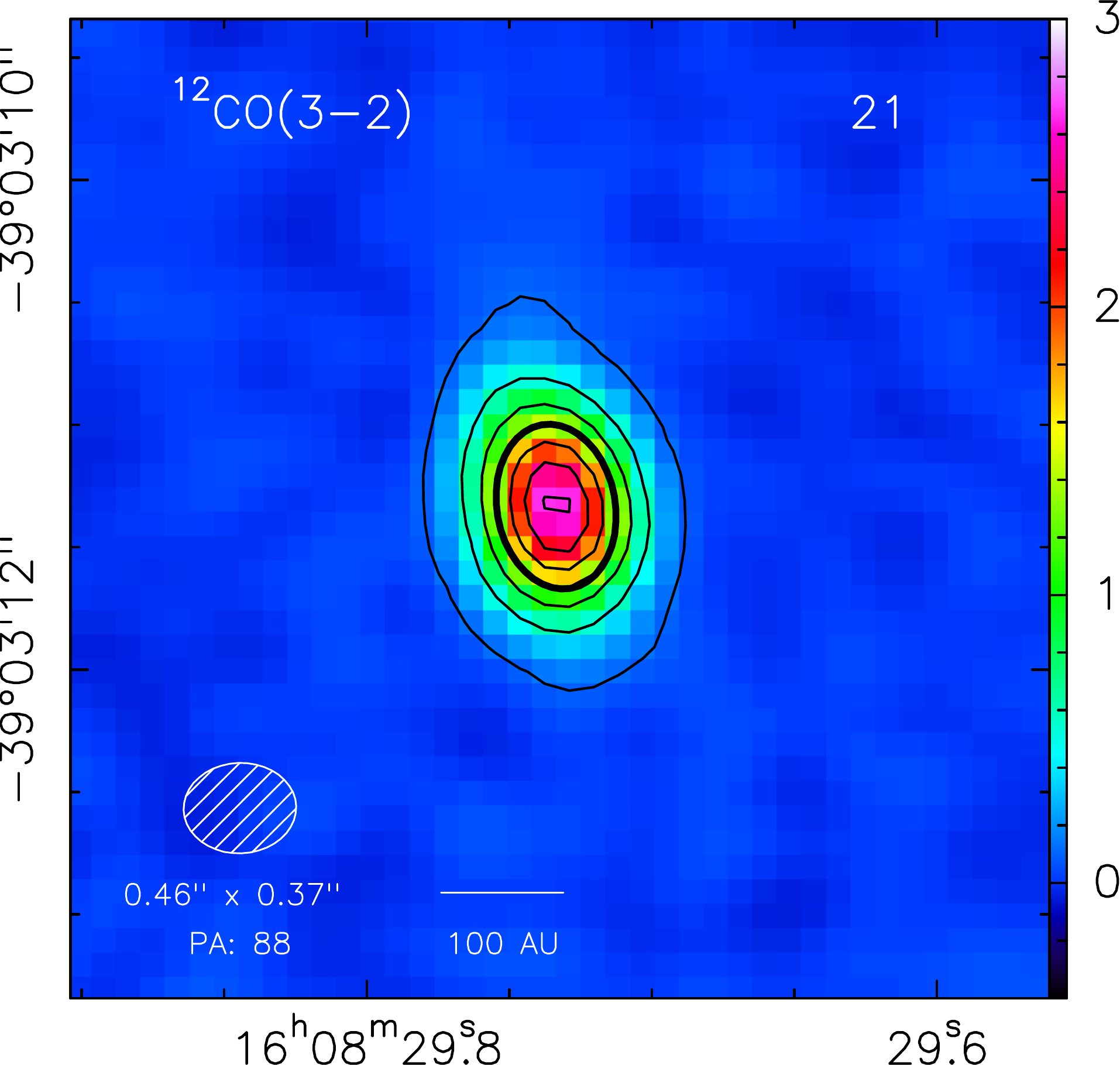}}
\hspace{0.3cm}\subfloat{\includegraphics[scale=0.32]{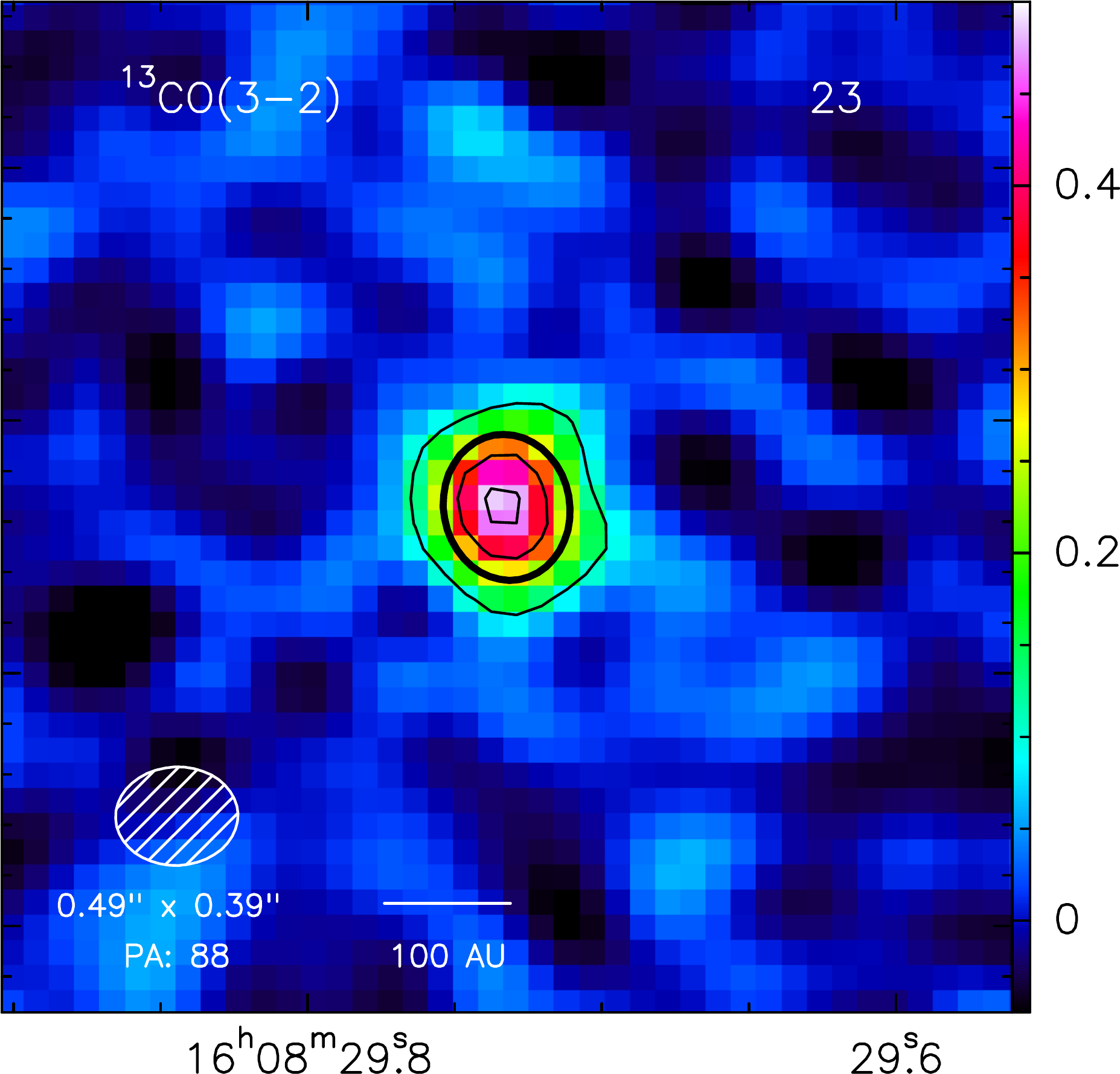}}
\hspace{0.3cm}\subfloat{\includegraphics[scale=0.32]{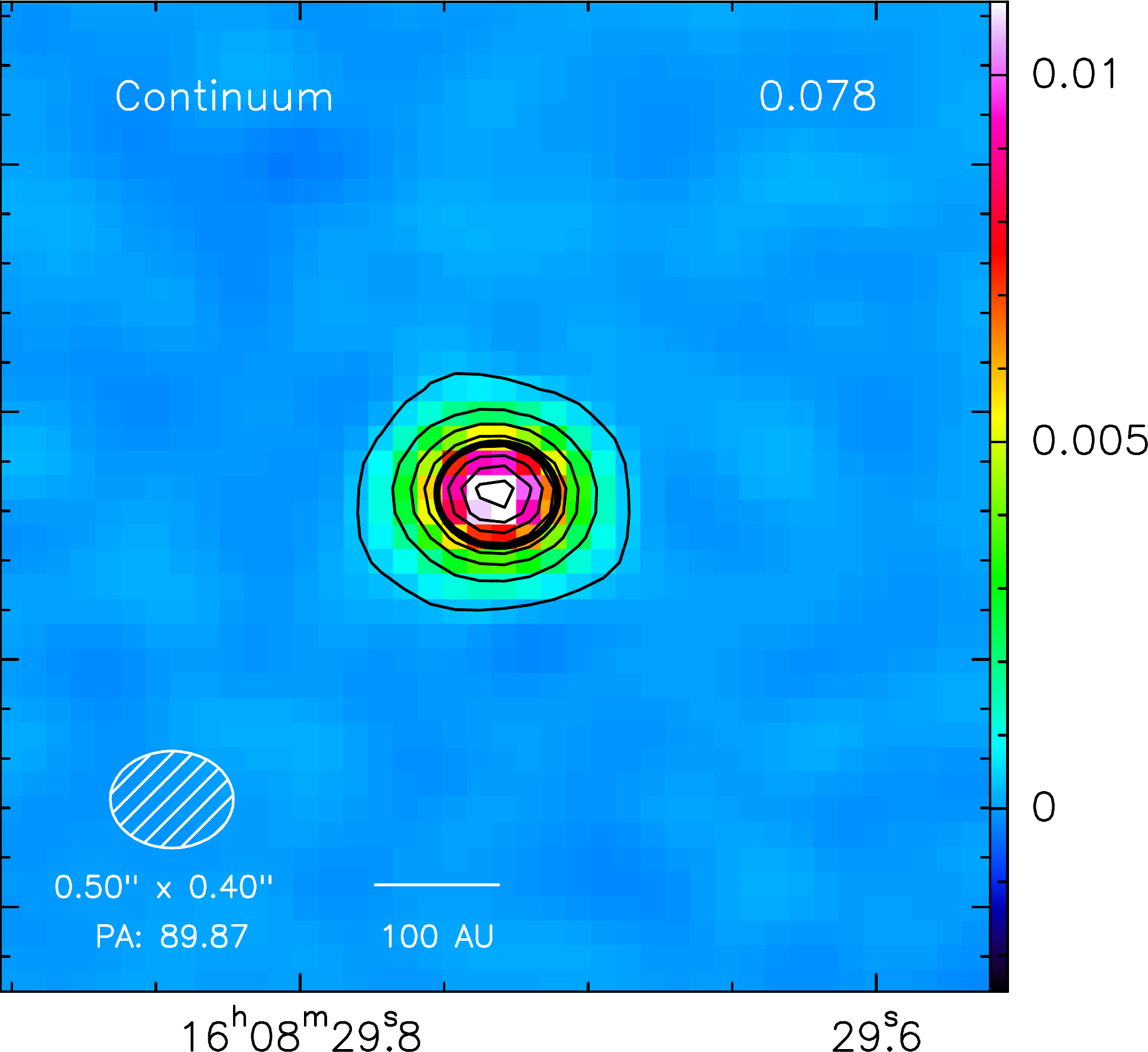}}
\caption{\textbf{Left:} moment zero of the $^{12}$CO(3-2) integrated from -15 to 20 km.s$^{-1}$. Contours start at 5\,$\sigma$ with 20\,$\sigma$ steps. \textbf{Middle:} moment zero of the $^{13}$CO(3-2) integrated from -15 to 20 km.s$^{-1}$. Contours start at 5\,$\sigma$ with 5\,$\sigma$ steps. \textbf{Right:} continuum map. Contours start at 5\,$\sigma$ with 20\,$\sigma$ steps. Noise level at 1\,$\sigma$ is indicated in the upper right corner of each panel; the unit is mJy/beam.km/s in the two left panels and mJy/beam in the right panel. The black ellipses on the three panels represent the FWHM intensity contours. White hatched ellipses represent the clean beam FWHM.}
\label{f:the3}
\end{figure*}

The interplay between accretion and ejection of matter is believed to be a crucial element in the formation of stars. Yet, the exact link between jets and accretion disks is still a critical issue in contemporary astrophysics \citep{ray07}. The most attractive possibility is a fundamental connection via angular momentum transfer from disk to jet by means of magnetocentrifugal forces, so that circumstellar material may continue to accrete onto the central object \citep[e.g.,][]{blandford82}. Exactly where this transfer occurs and how it impacts the disk physics is, however, still hotly debated \citep{ferreira06, pudritz07, shang07, romanova09, cabrit09}.
Owing to their low extinction and their proximity ($\approx$ 150 pc), accreting classical T Tauri stars (hereafter CTTSs) offer a unique laboratory to elucidate the connection between accretion and ejection.

A recent major breakthrough was the detection with the STIS instrument mounted on the HST of transverse Doppler shifts of 5 - 30\, km.s$^{-1}$ in optical and near-ultraviolet (hereafter NUV) emission lines on scales of 50 AU across the base of four close CTTS jets \citep{bacciotti02, woitas05, coffey04, coffey07}. 
If these shifts are due to jet rotation in a steady state outflow, they would imply that jets are magnetically launched from the disk surface at relatively small-scale radii of 0.1-1.6 AU, thus solving the long-standing problem of the jet origin in CTTS \citep{bacciotti02,anderson03,coffey04,woitas05}. However, effects other than rotation \citep[e.g., precession; see][]{cerqueira06} could cause transverse Doppler shifts similar to those detected in T Tauri jets. It is crucial that the underlying disks rotate in the same sense as the jet to support the jet rotation interpretation.
This comparison was conducted in three of the four systems mentioned above (\object{CW\,Tau}, \object{DG\,Tau}, and \object{RW\,Aur}), plus in \object{RY\,Tau}. In \object{RY Tau}, the study was inconclusive as no systematic Doppler shifts were found in the jet; see \cite{coffey15}. The sense of rotation of the jets of \object{CW\,Tau} and \object{DG\,Tau} \citep{coffey07}, traced by optical lines, coincides with the sense of rotation of their disks (\citealt{testi02}; Cabrit private comm., respectively). The jet rotation sense traced in near ultra-violet (NUV) emission lines at higher flow velocities is coherent for \object{DG\,Tau}, but these emission lines are below the detection threshold for \object{CW\,Tau} \citep{coffey07}. In RW\,Aur, the optical jet and disk rotations appear opposite (see \citealt{coffey04} and \citealt{cabrit06}). Here, jet rotation traced in the NUV was found to be compatible with the disk rotation sense (i.e., opposite to the optical jet rotation), but could only be measured in one lobe and at one of the two epochs \citep[spaced by six months; see ][]{coffey12}; because of its complexity and variability the authors declared \object{RW\,Aur} as unsuitable for jet rotation studies. We present the confrontation of jet and disk rotation senses for the fourth system, \object{Th\,28}. 

The T~Tauri star \object{ThA 15-28} \citep[hereafter \object{Th\,28};][]{the62}, also known as \object{Sz~102} \citep{schwartz77} or Krautter's star, is a young member of the \object{Lupus~3} association. Although uncertain, as the photospheric spectrum is veiled, a spectral type G8 to K2 was quoted for the central object of \object{Th\,28} \citep{graham88,hughes94,mortier11}. Even for a K star, the stellar emission from \object{Th\,28} is significantly underluminous: 0.03 L$_{\odot}$ \citep{mortier11}, which suggests that the disk is observed close to edge-on and obscures the central star. The derived disk luminosity, inferred from the spectral energy distribution (SED) integrated excess above the stellar photosphere, is $\sim$0.13 L$_{\odot}$ \citep{mortier11, merin08}. We adopt the recent value of d=185$^{\rm +11}_{\rm -10}$\,pc derived by \citet{Galli13} for the Lupus~3 cloud, with an improved convergent point search method, for the distance to the source.

\object{Th\,28} drives a bright and striking bipolar jet  whose axis lies at PA=98$^{\circ}$ \citep{krautter86,graham88}. The large proper motions derived for the distant knots are compatible with a close to edge-on geometry \citep[e.g.,][]{comeron10}. 
\object{Th\,28} is considered one of the best case for jet rotation detection in the HST sample, as transverse velocity gradients are well detected in both lobes of its optical atomic jet and are consistent in different optical lines \citep{coffey04}. In optical, the jet rotation sense is therefore well determined with a clockwise direction looking down the blueshifted jet lobe toward the source. The situation for NUV lines were inconclusive \citep{coffey07}. Only one measurement on the receding jet, that is, however, consistent with the sense
of rotation derived in optical, exceeded the incertitude level. 

\smallskip
In this paper, we report ALMA band 7 CO and continuum observations at 0.4$^{\prime\prime}$-0.5$^{\prime\prime}$ angular resolution of the \object{Th\,28} system,  aimed to detect the disk and determine its sense of rotation. We detail our observations and data reduction in Sect.~\ref{s:obs}, and demonstrate that the CO and continuum emissions clearly trace a disk in Keplerian rotation around \object{Th\,28} in Sect.~\ref{s:results}. The Sect.~\ref{s:analysis} offers a detailed analysis of the disk properties. In Sect.~\ref{s:discu} we discuss the implications for jet launching models and central source properties. We summarize our conclusions in Sect.~\ref{s:concl}.

\section{Observations and data reduction}
\label{s:obs}
\begin{table*}[htbp!]
\begin{center}
\addtolength{\tabcolsep}{+4pt}
\caption{Main observational parameters}
\label{t:paralma}
\begin{tabular}{lccc}
\hline
\hline
Parameter                  & $^{12}$CO                                             & $^{13}$CO                                                & Continuum                                             \\
\hline                                                                                                                                                                                                                                      
Frequency                  & 345.8 GHz                                             & 330.6 GHz                                                & 332.0$^{a}$ GHz                                       \\ 
Bandwidth                  & 469 MHz                                               & 469 MHz                                                  & 2 GHz                                                 \\
Native channel width       & 122 kHz                                               & 122 kHz                                                  & 15625 kHz                                             \\
Primary beam               & 18.2\arcsec                                           & 19.0\arcsec                                              & 18.9\arcsec                                           \\
Synthesized beam           & $0\farcs46\times0\farcs37$                            & $0\farcs49\times0\farcs39$                               & $0\farcs50\times0\farcs39$                            \\
Beam position angle        & 87.5$^{\circ}$                                        & 87.8$^{\circ}$                                           & 90.0$^{\circ}$                                        \\ 
$1\sigma$ rms$^b$          & $\sim$21\,mJy/beam.km/s                               & $\sim$23\,mJy/beam.km/s                                  & $\sim$0.078\,mJy/beam                                 \\
\hline
\end{tabular}
\end{center}
($^a$): The mean frequency was calculated assuming a S($\nu$)\,$\propto \nu^{-3}$ emission spectra accurately describes the ISM SED slope in the 2\,GHz spectral window. \\
($^b$): The rms presented here report 1$\sigma$ rms of the integrated maps shown in Fig.~\ref{f:the3}.
\end{table*}

The characteristics of our lines and continuum observations are detailed below. The resulting beam sizes and sensitivities are summarized in Table~\ref{t:paralma}.

The $^{12}$CO($J$=3$\rightarrow$2) and $^{13}$CO($J$=3$\rightarrow$2) emission lines plus the continuum emission of \object{Th\,28} were observed using ALMA Band 7 (275-373\,GHz). The observations were performed on April 28$^{}$ 2014 in two tracks of $\sim$\,40 minutes each during the ALMA cycle 1 campaign. The Band 7 data contained two spectral windows of 468.75\,MHz bandwidth each (3840 channels) that were tuned at 345.818\,GHz and 330.609\,GHz. This allowed us to simultaneously cover the $^{12}$CO($J$=3$\rightarrow$2) and $^{13}$CO($J$=3$\rightarrow$2) at 345.796\,GHz and 330.588\,GHz, respectively. A third spectral window of 2\,GHz bandwidth in 128 channels centered at 332.014\,GHz was dedicated to the detection of the continuum emission from the \object{Th\,28} dust disk. 
The phase center of the observations was taken to be the position of \object{Th\,28} as seen with HST: $\alpha$($J$2000) = 16$^{h}$08$^{m}$29$\overset{s}{.}$72 and $\delta$($J$2000) = -39$^{\circ}$03$^{\prime}$11$\overset{\prime\prime}{.}$01. 
The data were taken using the cycle 1 semiextended configuration of ALMA with baselines ranging from 20\,m to 480\,m. 

The data were reduced using the common astronomy software application \citep[hereafter CASA; see][]{mcmullin06}. We performed an initial correction for rapid atmospheric variations at each antenna using water vapor radiometer data and corrected for the time and frequency dependence of the system temperatures. Bandpass calibration was performed on the quasars \object{J1733-1304} and \object{J1517-2422}. The primary flux calibration was made using \object{Titan} and \object{J1517-2422}. Quasar \object{J1534-3526} was used in both observations to calibrate the time variation of the complex gains. Based on the dispersion between the fluxes derived for the phase calibrator in each observing session (using the two different amplitude calibrator), we estimate the absolute flux calibration to be accurate within $\sim$10\%. Owing to residual inconsistency in the phase calibration between the two tracks, we used the continuum spectral window of one track to further phase calibrate all spectral windows.
Imaging was carried out using the CLEAN algorithm of CASA. 
We used the Briggs weighting with a robust parameter of 0.5\footnote{Therefore giving slightly more weight to the longest baselines than with natural weighting.}. The channel spacing of 0.122\,MHz resulted in a maximum spectral resolution of $\sim$0.11\,km.s$^{-1}$ that we degraded down to 0.4\,km.s$^{-1}$ to detect weak $^{12}$CO and $^{13}$CO emission. The resulting root mean square (RMS) noise level is of 4 mJy/beam per 0.4\,km.s$^{-1}$ channel in $^{12}$CO(3-2) and $^{13}$CO(3-2). The continuum displays a RMS noise of 0.08 mJy/beam over its 2 GHz bandwidth. Panels (a), (b), and (c) in Figure~\ref{f:the3} show the $^{12}$CO(3-2), $^{13}$CO(3-2), and continuum, respectively, which will be described in Sect~\ref{s:results}.

\section{Results}
\label{s:results}

\begin{figure*}[ht!]
\centerline{
\subfloat{\includegraphics[scale=0.48]{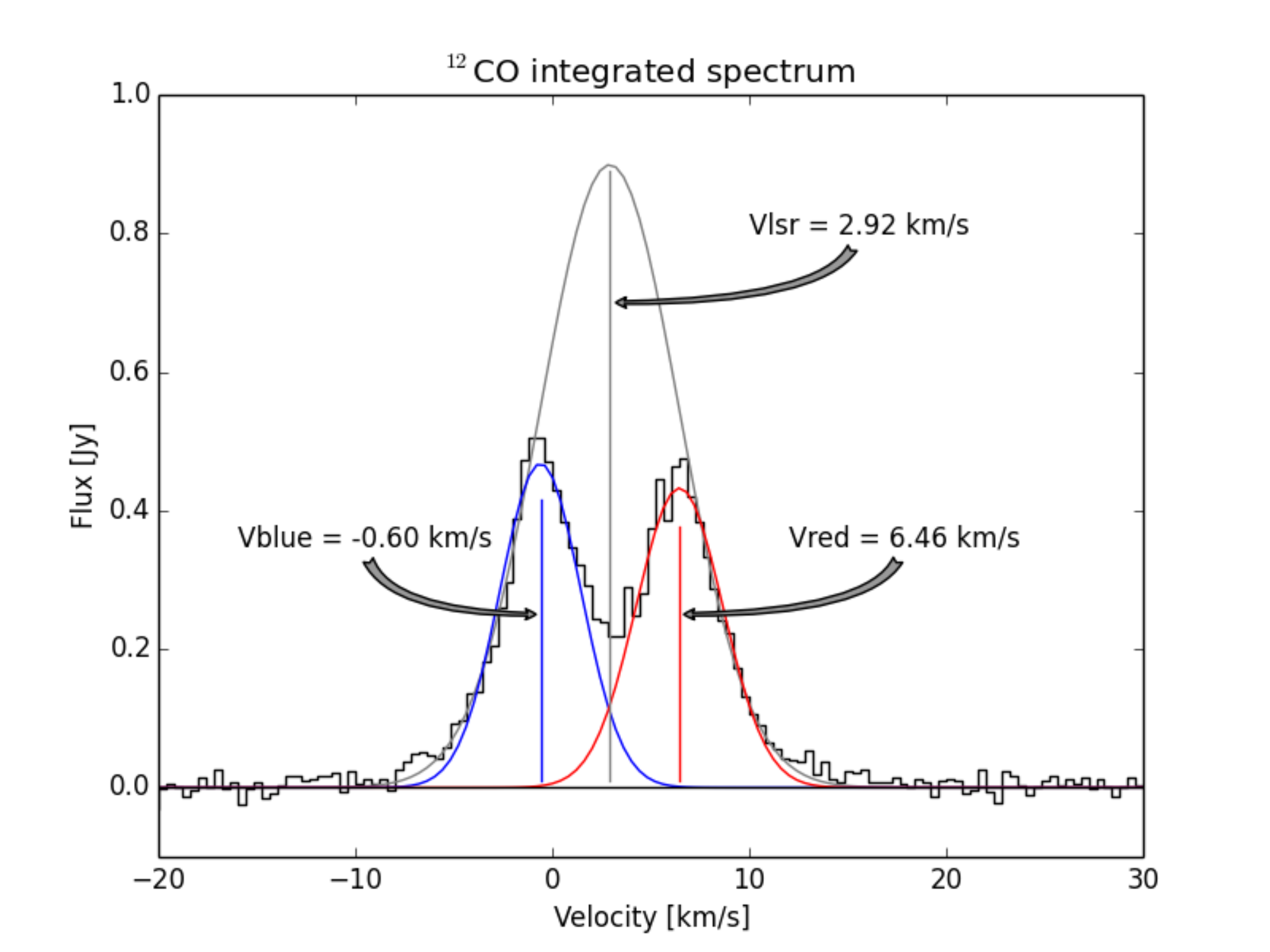}}
\hspace{-0.5cm}
\subfloat{\includegraphics[scale=0.48]{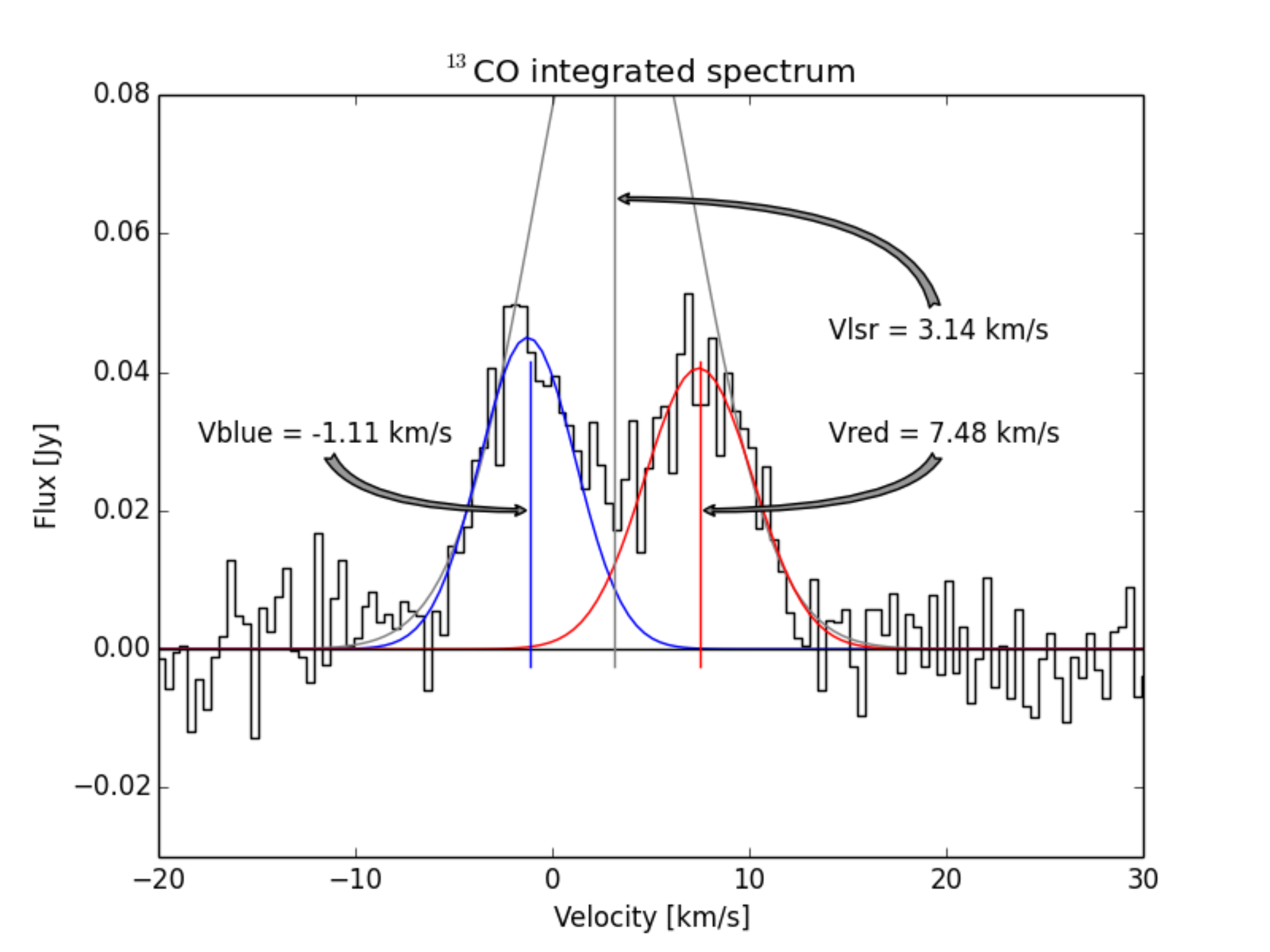}}\\}
\centerline{
\subfloat{\includegraphics[scale=0.38]{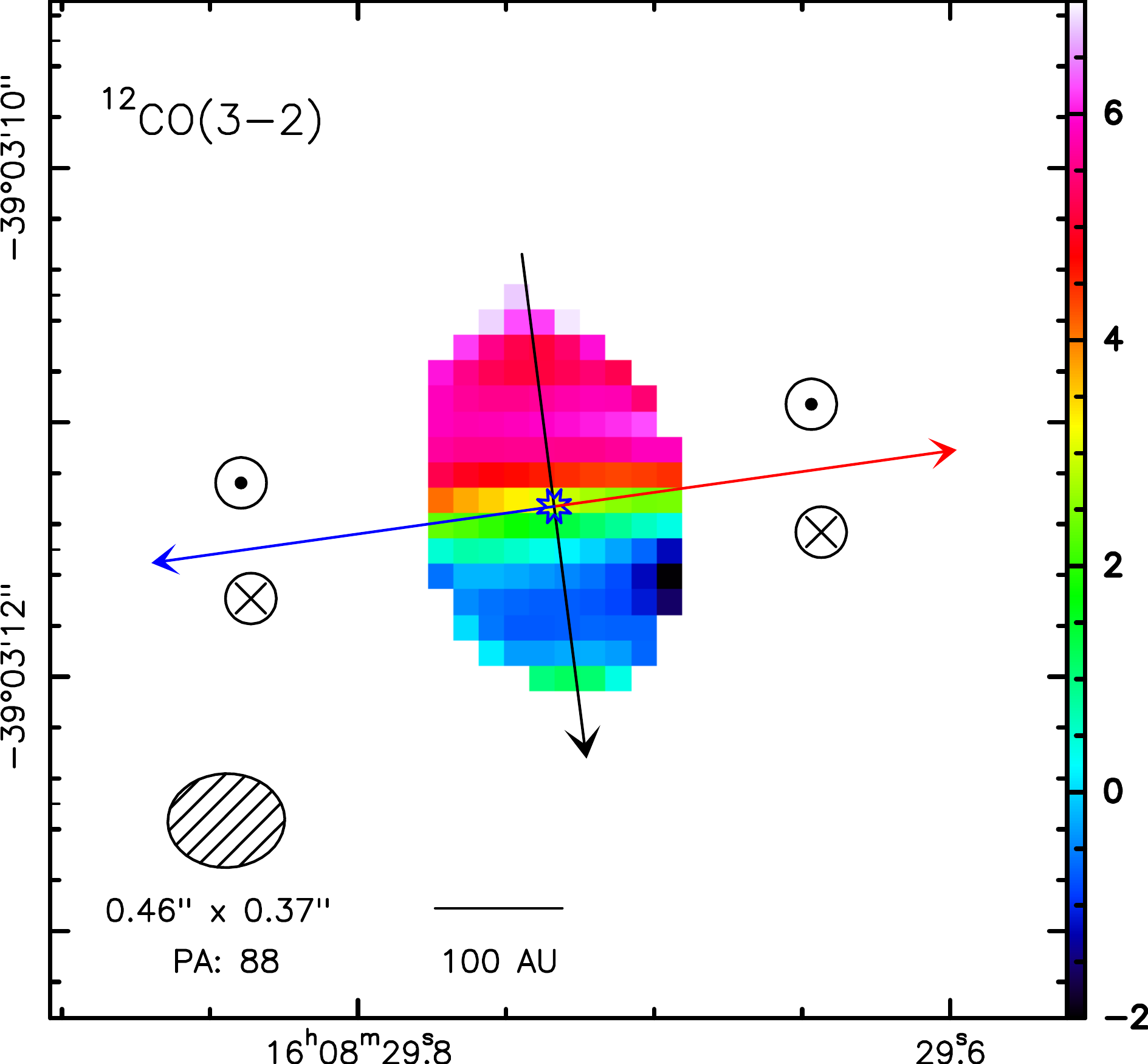}}
\hspace{+2cm}
\subfloat{\includegraphics[scale=0.38]{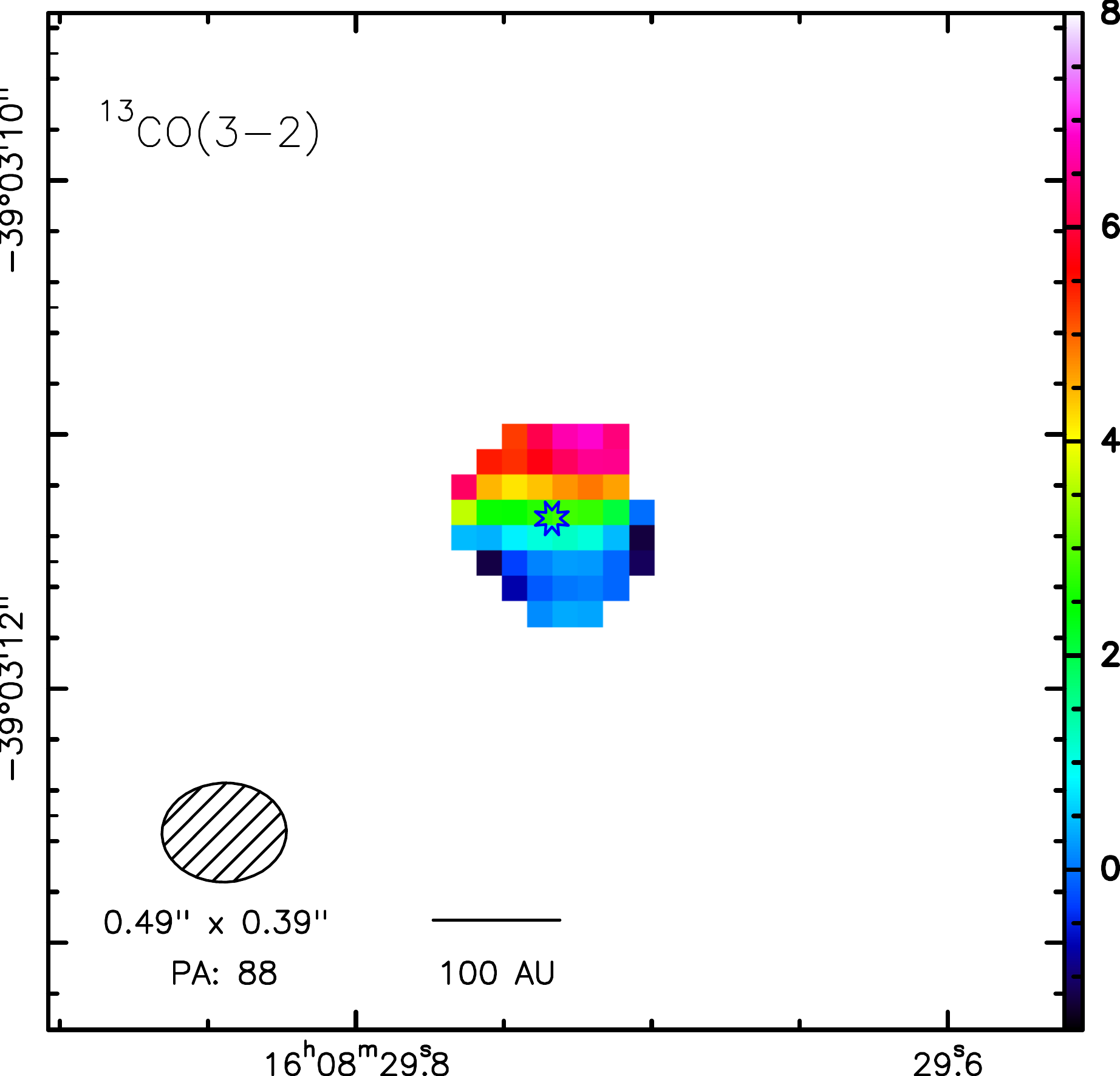}}}
\caption{\textbf{Top left}:  integrated spectra of the $^{12}$CO(3--2) emission line. \textbf{Top right: }integrated spectra of the $^{13}$CO(3--2) emission line. The integration surfaces correspond to areas where emission is above 5$\sigma$ on the corresponding integrated maps presented in Figure~\ref{f:the3}. \textbf{Bottom left:} first moment of the $^{12}$CO(3--2) emission line. The black arrow indicates the PA of the disk as seen in $^{12}$CO. The latter is used to build the PV diagram shown in Figure~\ref{f:pvd}. The red and blue arrows represent the jet axis of \object{Th\,28}, while the line-of-sight vectors represent their rotation sense in optical tracers derived by \citet{coffey04}. \textbf{Bottom right:} first moment of the $^{13}$CO(3--2) emission line. In these two first-moment maps, a clear velocity gradient due to the rotation of the circumstellar disk of \object{Th\,28} is seen.} 
\label{f:kine}
\end{figure*}

\subsection{CO emissions probing a disk in Keplerian rotation}
\label{ss:kepler}

Left panels of Figure~\ref{f:the3} show CO integrated emission maps over the velocity interval [-15 km.s$^{-1}$, 20 km.s$^{-1}$]. Both CO lines are clearly detected and centered at the source position with an integrated flux of 5.1\,Jy.km.s$^{-1}$ in $^{12}$CO and 0.60 Jy.km.s$^{-1}$ in $^{13}$CO. The top panel of Fig.~\ref{f:kine} shows the integrated spectra of the $^{12}$CO(3--2) and of the $^{13}$CO(3--2) inside the area defined by the 5$\sigma$ contours of Figure~\ref{f:the3} (left) and Figure~\ref{f:the3} (middle), respectively. The morphology and kinematics of the CO emission show clear signatures of a protoplanetary disk in Keplerian rotation:
\begin{itemize}
\item[\textit{(a)}] The emission is compact ($<$200\, AU) and elongated in $^{12}$CO.
\item[\textit{(b)}] The elongation is perpendicular to the atomic jet position angle (PA).  
\item[\textit{(c)}] The CO lines show the double-peaked shape characteristic of a disk in Keplerian rotation.
\item[\textit{(d)}] The position-velocity diagram along the PA of the disk is also compatible with a disk in Keplerian rotation.
\end{itemize}
We detail each of these items below.

\paragraph{\textit{(a)} Compact and elongated CO emission: }
The $^{12}$CO emission shows an elliptical shape with a main axis close to the north-south direction, roughly perpendicular to the main axis of the beam. It is therefore clearly resolved in the north-south direction. 
Indeed, the emitting area with fluxes above half maximum emission is an ellipse of 0.68$\arcsec\times$0.48$\arcsec$ at PA$\approx$10$^{\circ}$ (black ellipse on Fig.~\ref{f:the3}, left) while the beam is 0.46$\arcsec\times$0.37$\arcsec$ at PA$\approx$88$^{\circ}$.
The $^{13}$CO displays a more compact emission than the $^{12}$CO. The area of half maximum emission corresponds to an ellipse of 0.58$\arcsec\times$0.50$\arcsec$ at PA $\approx$10$^{\circ}$ (see black ellipse on Fig.~\ref{f:the3}, middle). When compared to the beam size, 0.49$\arcsec\times$0.39$\arcsec$ at PA$\approx$88$^{\circ}$, the $^{13}$CO appears only marginally resolved in the north-south direction. The compact scale of the emission argues against an origin in an infalling envelope. The CO emission therefore most likely traces the accretion disk around \object{Th\,28}.

\paragraph{\textit{(b)} CO emission perpendicular to the PA of the atomic jet: }
To derive the intrinsic size and geometry of the CO emission, we performed 2D-Gaussian fits directly in the uv plane with the UV-FIT procedure of GILDAS\footnote{See the following webpage for details: \href{https://www.iram.fr/IRAMFR/GILDAS/}{IRAM}}. The result of these fits are reported in Table~\ref{t:parfit}. The 2D-Gaussian fits have FWHM of 0.82$\pm0.01\arcsec$ $\times$ 0.24$\pm0.01\arcsec$ at a PA=7.3$\pm0.4^\circ$ in $^{12}$CO and 0.66$\pm$0.03$\arcsec$ $\times$ 0.19$\pm$0.09$\arcsec$ at PA=7.6$\pm$3.7$^{\circ}$ in $^{13}$CO.

The PA of the long axis of the CO emissions (see (a) and (b)) are therefore perpendicular, within the uncertainties, to the PA of the optical blueshifted jet (98$^{\circ}$). 

\paragraph{\textit{(c)} Double-peaked profile of the CO lines: }
The integrated CO line profiles are double peaked with symmetric wings (see Fig.~\ref{f:kine}-Top), exhibiting a strong resemblance to the typical profile shape of rotating Keplerian disks in CTTS \citep{duvert00,guilloteau94}. The $^{12}$CO and $^{13}$CO show peak fluxes of $\sim$0.45\,Jy and 0.04\,Jy, respectively. The apparent separations of the two velocity peaks toward \object{Th\,28} is larger in $^{13}$CO than in $^{12}$CO: $\Delta$v = 8.7$\pm 0.1$\,km.s$^{-1}$ and 7.1$\pm$0.3\,km.s$^{-1}$, respectively. Such a behavior is expected for a Keplerian disk, since the intensity peaks roughly correspond to the outer radius of the disk, which should be larger in the more optically thick $^{12}$CO(2-1) line.

\begin{figure}[htb!]
\includegraphics[scale=0.32]{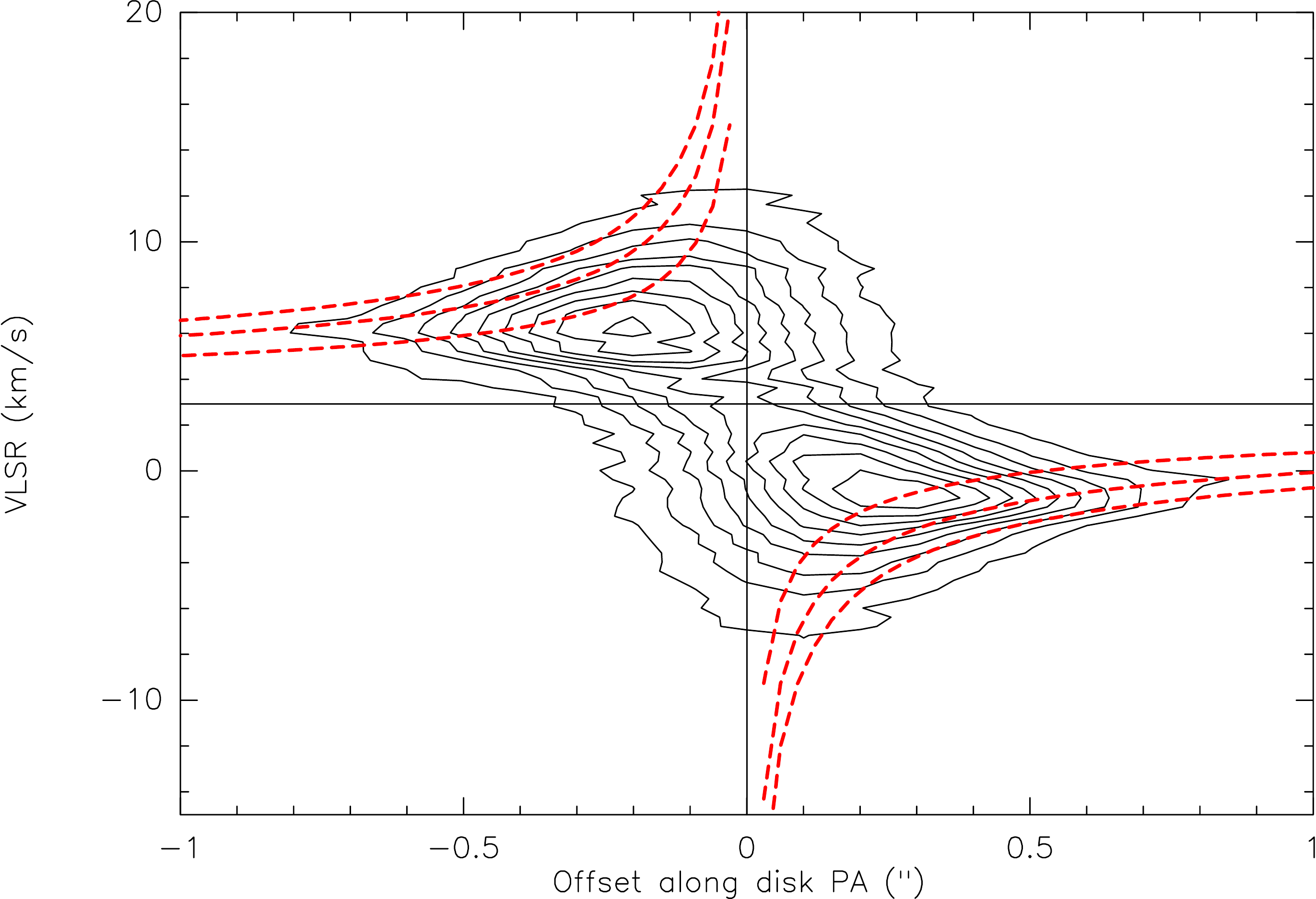}
\caption{PV diagram of the $^{12}$CO emission along the disk PA (see black arrow on Figure~\ref{f:kine}, bottom left). Contour star at 32\,mJy/beam with 32\,mJy/beam steps. The plot is superimposed with expected Keplerian motions in a disk inclined by 80$^{\circ}$ for a central object of 1, 2, and 3\,M$_\odot$ (internal, central, and external curves respectively).}
\label{f:pvd}
\end{figure}

\paragraph{\textit{(d)} The position-velocity (PV) diagram}of the $^{12}$CO emission line along the PA=7.3$^{\circ}$ of the disk is shown in Fig.~\ref{f:pvd}. It further confirms the hypothesis of a disk in rotation with a quasi-perfect point symmetry both in position and velocity. Such a pattern is indeed indicative of a globally axisymmetric rotating structure around \object{Th\,28}. We have superimposed Keplerian curves for stellar central masses M$_\star$=1, 2, and 3\,M$_\odot$ (see Sect.~\ref{ss:star} for a more detailed analysis), assuming the disk has an inclination of 73 degrees.

\smallskip

With the supporting evidence presented above, we claim to observe a protoplanetary disk in Keplerian rotation around \object{Th\,28}. 

\begin{table*}[htbp!]
\begin{center}
\addtolength{\tabcolsep}{+1pt}
\caption{Fit to the observable features}
\label{t:parfit}
\begin{tabular}{l|c|cccc|c|c}
\hline
\hline
Parameter      & Flux$^a$               & \multicolumn{4}{c}{Elliptical Gaussian fit in the uv plane$^{b}$}         & $V_{\rm lsr}^{c}$           & Peak separation        \\
               &                        & Major axis          & Minor axis          & PA            & Inclination   & -                           & $\Delta$v              \\
               & Jy                     & [\arcsec]           & [\arcsec]           & [$^{\circ}$]  & [$^{\circ}$]  & [km/s]                      & [km/s]                 \\
\hline                                                                                                                                                                                                                   
$^{12}$CO      & 5.10                   & 0.82$\pm$0.01       & 0.24$\pm$0.01       & 7.3$\pm$0.4   & 73$\pm$1      &  2.92$\pm$0.05              & 7.1$\pm$0.1            \\
$^{13}$CO      & 0.60                   & 0.67$\pm$0.03       & 0.19$\pm$0.09       & 7.6$\pm$3.7   & 73$\pm$9      &  3.14$\pm$0.10              & 8.7$\pm$0.3            \\
Continuum      & 13.3$\times 10^{-3}$   & 0.26$\pm$0.01       & 0.09$\pm$0.04       & 15$\pm$4      & 69$\pm$11     &   -                         &  -                     \\
\hline     
\end{tabular}
\end{center}
($^{a}$): Integrated fluxes inside the area defined by a 5$\sigma$ contour. The emission lines are integrated in the -15 to 20 km.s$^{-1}$ range. \\
($^{b}$): The fits in the uv plane are centered at the following coordinates, $\alpha$($J$2000)=16$^{h}$08$^{m}$29$\overset{s}{.}$73 and $\delta$($J$2000)=-39$^{\circ}$03$^{\prime}$11$\overset{\prime\prime}{.}$33  \\
($^{c}$): The central velocities are determined from the fit of a single Gaussian to the wings of the spectra (see Fig.~\ref{f:kine}-Top) \\
\end{table*}

\subsection{Continuum emission of the disk}
\label{ss:cont}
The Figure~\ref{f:the3} (right) shows the 0.9\,mm continuum emission of \object{Th\,28}, centrally peaked at the position of the source. With a signal-to-noise (S/N) ratio of $\sim$150, we report the unambiguous detection of continuum emission in the circumstellar disk of \object{Th\,28}. The comparison of the beam (0.50\arcsec$\times$0.40\arcsec with PA of 88.8$^{\circ}$) to the emission area with fluxes above half of the peak flux (0.50\arcsec$\times$0.41\arcsec with PA of 88.0$^{\circ}$) indicates unresolved continuum emission from \object{Th\,28} at 0.90\,mm. The mm-dust emission is therefore compact, in accordance with an emission originating from a disk.

As for the line results (see Sect.~\ref{ss:kepler}), to free ourselves of deconvolution uncertainties, we performed a fit of the disk parameters in the uv plane. Keeping in mind that the continuum appears unresolved in the image plane, it reveals a disk of $\sim$0.26\arcsec$\times$0.10\arcsec at PA=15$\pm4^{\circ}$ (see Table~\ref{t:parfit}), which is roughly compatible with those derived in $^{12}$CO \& $^{13}$CO.

\paragraph{Mass of the disk:} 

If we assume optically thin emission from dust grains at $T$=20 K, a dust opacity coefficient $k_{\nu} = 0.1 {\rm cm^2.g^{-1}}\times(\nu/1000\rm GHz)^\beta$ \citep[very similar to that advocated by][including a gas to dust ratio of 100 by mass]{hildebrand83} with $\beta \sim$ 0.5 \citep{beckwith91}, our total flux density of 13.3\,mJy implies a total mass of $\sim$8.5$\times10^{-4}$\,M$_{\odot}$ at the distance of 185\,pc of \object{Th\,28}. This circumstellar disk mass is relatively low but is not inconsistent with known CTTS disk masses \citep[4$\times10^{-3}$\,M$_\odot$ with a 0.72 dex standard deviation;][]{andrews05}. 
Also, this value is highly sensible to the assumptions made on the parameters listed above. For instance a variation of $-$20\% (resp. +20\%) of the dust temperature implies an augmentation (resp. diminution) of the disk total mass to 1.2$\times10^{-3}$\,M$_\odot$ (resp. 6.6$\times10^{-4}$\,M$_\odot$), while varying the opacity index $\beta$ by minus or plus 20\% changes the total disk mass to 7.6$\times10^{-4}$\,M$_\odot$ and 9.4$\times10^{-4}$\,M$_\odot$, respectively.


\section{Analysis}
\label{s:analysis}

\subsection{$V_{\rm lsr}$ of the source}
\label{ss:vlsr}
To determine the $V_{\rm lsr}$ of \object{Th\,28}, we fitted a single Gaussian component to the high-velocity wings of the $^{12}$CO and $^{13}$CO profiles (see top of Fig.~\ref{f:kine} and Table~\ref{t:parfit}). We derived a central velocity of $V_{\rm lsr}\approx2.92$\,km.s$^{-1}$ in $^{12}$CO and $V_{\rm lsr}\approx3.14$\,km.s$^{-1}$ in $^{13}$CO. In the following, we adopt a $V_{\rm lsr}$=3$\pm0.1$\,km.s$^{-1}$ for \object{Th\,28}. This value is compatible with the range of $V_{\rm lsr}$ observed toward the Lupus~3 cloud \citep{james06}.

\subsection{Disk inclination }
\label{ss:diskincl}

The CO emission aspect ratio derived from the 2D-Gaussian fits in the uv plane
provide estimates of the inclination to the line of sight of 73$\pm$1$^{\circ}$ and 73$\pm$9$^{\circ}$ from the $^{12}$CO and $^{13}$CO emission lines, respectively (see Table~\ref{t:parfit}). This inclination is comparable but somewhat lower than the inclination of 82.3$\pm$2$^{\circ}$ for the optical jet determined from the proper motions of distant knots (see Appendix). However, the CO emission may be underestimating the true disk inclination. Indeed, CO emissions are expected to arise above the disk plane at typical altitudes of 2-3 disk scale heights \citep{dartois03}. For a close to edge-on geometry, the apparent ellipse in the plane of the sky has a larger semiminor axis than a flat disk, mimicking a less inclined flat disk.

\subsection{Disk rotation sense }
\label{ss:diskrot}

The lower panels of Figure~\ref{f:kine} show the first moment map of the $^{12}$CO(3--2) and $^{13}$CO(3--2) emission lines. A velocity gradient along the PA of the disk is striking from those plots. They reveal that the northeastern part of the disk is redshifted, while the southwestern part is blueshifted.
We therefore report unambiguous detection of the sense of rotation of the \object{Th\,28} disk. The derived sense of rotation is opposite that derived for the optical jet with the HST (see Figure~\ref{f:kine}). We discuss the implications of this result for jet launching models in the following section.

\subsection{$^{12}$CO versus $^{13}$CO }
\label{ss:versus}

To derive the overall flux ratio of the $^{13}$CO emission line with respect to the $^{12}$CO emission line, we integrated both lines \textit{(i)} in the velocity range -15\,km.s$^{-1}$ to 20\,km.s$^{-1}$ and \textit{(ii)} over the area where the $^{13}$CO displays fluxes above 5$\sigma$ (see Fig.~\ref{f:the3}-Middle). In so doing, we achieve integrated fluxes of $\sim$3.75\,Jy.km.s$^{-1}$ and $\sim$0.60\,Jy.km.s$^{-1}$ for the $^{12}$CO and $^{13}$CO, respectively. The integrated flux ratio, S($^{13}$CO)/S($^{12}$CO), is therefore of $\sim$0.16. 
The discrepancy between the integrated flux ratio (0.16) and the fractional abundance of CO isotopes ($\rm [^{13}CO]/[^{12}CO]$\,$\sim$0.017 in the ISM) implies the $^{12}$CO to be optically thick. Nevertheless, an optically thick $^{12}$CO line profile supplies a valuable indication on the temperature profile in the disk \citep[hereafter BS93]{beckwith93}; see Section~\ref{ss:temp}.

\subsection{Temperature profile (T(K) versus radius) in the disk }
\label{ss:temp}
Since the $^{12}$CO emission line is optically thick (see above), its spectrum is a direct probe of the temperature as a function of radius. Following the analysis developed by BS93, if the temperature profile follows a power law in radius ($r$), as $T\propto r^{-q}$, the flux density is roughly proportional to $v_{\rm obs}^{3q-5}$ in the high-velocity wings. Performing the fit of the flux density on the high-velocity blue wing [-8.0\,km.s$^{-1}$; -2.4\,km.s$^{-1}$] of the $^{12}$CO emission line, we find an index q=0.8$\pm$0.2. Under these assumptions, the temperature profile of \object{Th\,28} is steeper than the typical values derived in CTTS disks \citep[0.5-0.75; see][]{isella09,beckwith91}.


\section{Discussion}
\label{s:discu}

Our ALMA observations of \object{Th\,28} reveal, for the first time, the millimeter CO and continuum emission from its accretion disk. We first discuss the implications brought by these observations for jet rotation studies. We then briefly discuss additional first order constraints on the \object{Th\,28} system.

\subsection{Implications for jet rotation studies}

 Our high S/N ratio and well-behaved ALMA CO observations unambiguously establish that the \object{Th\,28} disk rotates in a sense opposite to that derived with the HST for both lobes of the \object{Th\,28} atomic optical jet by \citet{coffey04}. Now, consistency between disk and jet rotation has been investigated in five systems. One was not informative, since the jet rotation sense could not be detected (RY Tau). Two of them (\object{DG Tau}, \object{CW Tau}) show consistent disk and optical jet rotation sense, while the two others (\object{RW Aur} and now \object{Th\,28}) show inconsistent rotation sense. However, \object{RW\,Aur-A} is peculiar since its star shows periodic radial velocity variations possibly signaling a close companion \citep{petrov01}, which might confuse its jet dynamics, for example, through jet precession/wiggling or shock asymmetries \citep{cerqueira06}. On the contrary, the \object{Th\,28} jet was one of the best cases of jet rotation detection, with different optical lines and both lobes of the jet giving similar transverse velocity gradients. So, the discrepancy observed in that most favorable case between jet and disk rotation suggests broader implications for optical jet rotation studies.
We discuss below three possible explanations for the observed discrepancy:

1) The disk rotation is probed on radial distances r~$\geq$~20\,AU from the source, while the jet is expected to originate from much smaller disk radii, r~$<$~5~AU \citep{ray07}. One could imagine either a true change of rotation sense in the inner disk regions or
a warp between the inner and outer disk that could induce an artificial change of rotation sense in the outer disk due to projection effects. Although we cannot fully exclude these possibilities, they seem unlikely as both the inclination and the PA of the outer CO disk axis agree well with that of the atomic jet, indicating no significant change of disk geometry between AU to tens of AU scales. In addition, the high inclination of the outer disk rotation axis means that line-of-sight velocities are very close to the underlying disk rotation velocities and that projection effects should be minimized. Moreover, the disk of \object{Th\,28} does not show any evidence of perturbation in its PV diagram at high velocities that could betray such behavior.

2) The observed transverse velocity gradients in jets do trace rotation but in a nonstationary part of the flow. Indeed, shocks and/or nonstationary ejection may significantly alter the apparent rotation of the flow, either amplifying it \citep{fendt11} or even reversing it \citep{sauty12} owing to the exchange of angular momentum between the flow and the magnetic field. Also, \cite{staff15} recently presented 3D magnetohydrodynamic simulations of disk winds where the less opened magnetic field configuration results in a wide, two-component jet. They show that kink mode creates a narrow corkscrew-like jet with regions in which jet rotation at $\pm$2\,AU from its axis is opposite to the disk. However such counter-rotating jet signatures would remain unresolved at the current achievable angular resolution of 0.1$^{\prime\prime}$ (=14\,AU) with the HST in the optical domain.

 3) Optical transverse velocity gradients in jets do not trace rotation in the jet body but rather a departure from axisymmetry. Indeed, high angular resolution observations show evidence of small-scale jet axis wiggling and/or asymmetric shock fronts \citep[see, e.g.,][]{white14,coffey15}. \citet{white14} recently demonstrated in the \object{DG Tau} case that spurious rotation signatures on the same magnitude as those observed in the optical can be mimicked if the jet axis wiggling is not properly taken into account. Moreover, transverse velocity gradients traced by optical emission lines probe an intermediate velocity component (IVC) of the flow, the nature of which is still unclear. \citet{white14} proposed that the IVC in \object{DG\,Tau} traces a dragged mixing layer between a fast inner jet and an outer slower wind. Therefore, transverse gradients detected in the NUV domain, which probe a faster and more collimated component of the flow, may trace jet rotation more reliably. The NUV velocity gradients detected in two sources so far (\object{DG\,Tau} and \object{RW\,Aur}) appear to rotate in the same sense as their underlying disks. However, \object{RW\,Aur} shows a time-variable behavior that renders it unsuitable for jet rotation studies. In the case of \object{Th\,28} no sense of rotation could be reliably determined in the NUV. Difficulties were encountered 
in this plane-of-sky jet because of a large, low-velocity absorption feature that closely coincides with the jet radial velocity. Further studies are clearly needed to establish whether observed NUV velocity gradients do indeed trace jet rotation. 

In any case, the implications are that the velocity gradients derived from optical lines in T Tauri jets cannot be used to infer jet launching radius using the method proposed by \citet{bacciotti02,anderson03,ferreira06}. Indeed, application of this method requires that the flow is both stationary and axisymmetric. Both assumptions seem inconsistent with the discrepancy found between the optical jet and disk rotation sense in Th 28.

\subsection{Implications for the nature of Th\,28}
\label{ss:star}

\begin{figure*}[htb!]
\begin{center}
\includegraphics[scale=0.19]{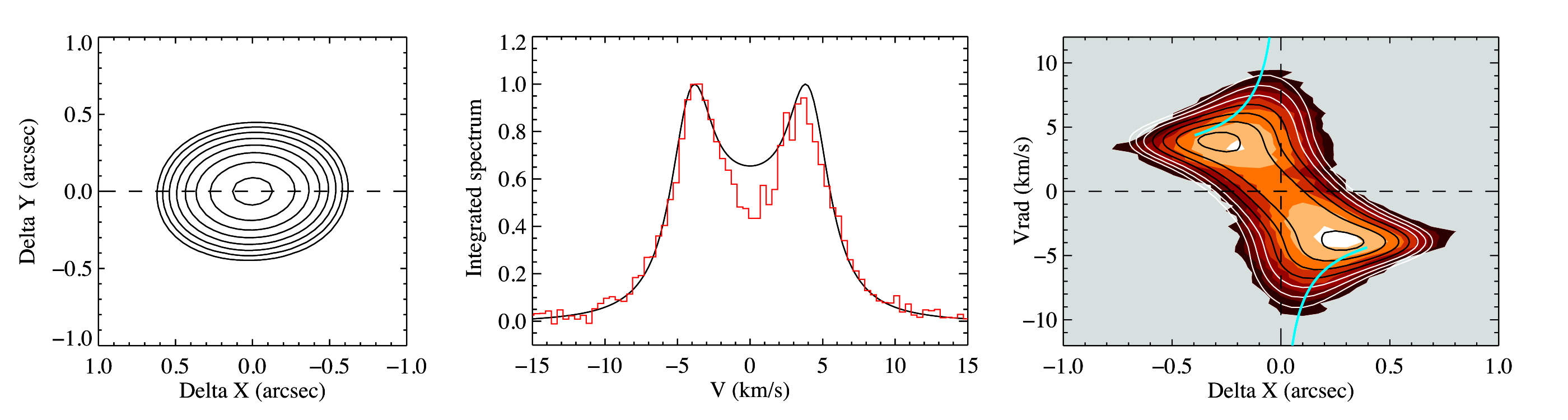}
\end{center}
\caption{Predictions of the $^{12}$CO emission for flat, optically thick disks viewed at 80$^\circ$ inclination. The model corresponds to a solution with R$_{\rm out}$=92.5\,AU, M$_\star$=1.6\,M$_{\odot}$. The model was convolved with a beam of 0.46$^{\prime\prime}$ by 0.37$^{\prime\prime}$ with a PA=88$^{\circ}$ .The left panel shows the predicted $^{12}$CO integrated intensity map; the middle panel shows the $^{12}$CO integrated spectra (data en red and model in black); and the right panel shows the predicted position-velocity diagram along the disk PA (black and white contours) overlaid on observations (background color map). The blue curve shows the theoretical projected Keplerian velocity profile. In all maps, contours start at 90\% of the peak and decrease by $\sqrt{2}$.}
\label{fig:pvmod}
\end{figure*}

\cite{comeron10} derived a central stellar mass of 0.6-0.9 M$_{\odot}$ from the maximum velocity of 450\,km.s$^{-1}$ observed in the redshifted wings of the H$\alpha$ emission line profile, assuming that this velocity traces the free-fall velocity of infalling gas at the base of the accretion columns. In the following, we argue that the combination of the spatial extension and peak velocity separation observed in $^{12}$CO  indicates a stellar mass more in the range 1-2\,M$_{\odot}$. 

An accurate derivation of the central stellar mass would require a detailed disk modeling, with radiative transfer effects properly taken into account, and fits of the spectral energy distribution. Such a detailed analysis goes beyond the scope of the present paper. Instead, we propose a simple derivation of the central stellar mass standing on the $^{12}$CO emission line. To meaningfully constrain the central stellar mass from the position-velocity diagram, it is critical to take projection and beam convolution effects into account.
To do so, we build a simple geometrical model of an optically thick flat disk with a radial power law temperature
distribution of index $q$ and we convolve this model with the effective ALMA beam of the $^{12}$CO observations. 
We fix the disk inclination to i=80$^{\circ}$ and the power law index $q$ of the temperature profile to q=0.8, which is 
most compatible with the shape of the wings in the $^{12}$CO profile (see Sect.~ \ref{ss:temp}). In order to reproduce the observed $^{12}$CO brightness profile along the disk major axis, a disk outer radius of R$_{\rm out}$=0.5$^{\prime\prime}$ is required, corresponding to 92.5\,AU at the adopted 
distance of d=185\,pc for \object{Th\,28}. The observed velocity separation between the peaks is then best fitted with a stellar mass 
of 1.6\,M$_{\odot}$. The model described above depends on the index of the temperature profile, $q$, whose value is uncertain by 25\,\% (see Sect.~\ref{ss:temp}). Spreading this uncertainty in the model, we derive an uncertainty on the mass of the central star of $\pm$0.2\,M$_\odot$. Moreover, a central stellar mass in the range 1-2\,M$_{\odot}$  appears more compatible
with a spectral type of early K as quoted in the literature for this object. Using the pre-main-sequence tracks of \citet{siess00}, a spectral type of K2 (T$_{eff}$= 4900-5000 K) corresponds to M$_{\rm star}\ge$ 2\,M$_{\odot}$ for ages less than 3.5\,Myrs.

Concerning the low luminosity of the central object, our best fit to the CO emission profiles gives V$_{\rm lsr}$=3$\pm0.1$ km.s$^{-1}$, which is consistent with the range of values observed in the Lupus~3 cloud \citep{james06}. It is then unlikely that the low luminosity of the central \object{Th\,28} star could be related to a larger distance. 
One would expect before extinction a $\sim$1.9\,L$_\odot$ luminosity from a 1.5\,M$_\odot$ protostar of age $\le$ 3.5 Myr \citep{siess00}\footnote{see also \href{http://www.astro.ulb.ac.be/~siess/pmwiki/pmwiki.php/WWWTools/PMS}{Webpage Siess}}. Therefore an extinction of 4.5\,mag would be sufficient to attenuate the star down to the luminosity observed for \object{Th\,28} \citep[0.03\,L$_\odot$, see][]{mortier11}. Such an extinction appears easily achievable from a close to edge-on system and our ALMA observations confirm a large inclination to the line of sight (i $\ge$ 73$^{\circ}$). This is also consistent with the fast proper motion of the jet knots at the distance of Lupus~3 when compared to their radial velocities (see Sect.~\ref{ss:diskincl}).

\section{Conclusions}
\label{s:concl} 

We have observed the T Tauri star \object{Th\,28} during the ALMA-cycle 1 campaign in Band 7 (275-373\,GHz). We detected $^{12}$CO($J$=3$\rightarrow$2), $^{13}$CO($J$=3$\rightarrow$2) and continuum signatures of a Keplerian accretion disk around \object{Th\,28}. 

\begin{itemize}

\item[$\bullet$] The $^{12}$CO emission is clearly resolved and elongated along PA=7.3$^{\circ}$. The morphology of the disk seen in CO matches very well the morphology derived for the large-scale atomic jet. The PA of the disk is perpendicular to that of the jet, and both inclinations are comparable.

\item[$\bullet$]  The $^{12}$CO shows kinematics that are consistent with a disk in Keplerian rotation. Indeed its double-peak integrated profile, plus its PV diagram along the PA of the disk, are characteristic of Keplerian rotation.

\item[$\bullet$] We derive a V$_{\rm lsr}$=3$\pm$0.1 km.s$^{-1}$ for the central source in agreement with the range of values observed in the Lupus~3 cloud.  

\item[$\bullet$] We constrain the power law index, $q$, of the temperature distribution to $q\simeq$0.8 and the R$_{\rm out}$ to $\simeq$ 95\,AU.

\item[$\bullet$]The combination of large R$_{\rm out}$ and peak velocity separation in $^{12}$CO of $\Delta$v=7\,km.s$^{-1}$ is best reproduced with a central stellar mass of 1.4-1.8\,M$_\odot$. This is also consistent with early K spectral type estimates for this source.

\item[$\bullet$]The rotation sense of the disk is well detected with our ALMA CO observations and this direction is opposite that of the transverse velocity shifts previously detected with HST in the optical jet.
This discrepancy in rotation senses implies that velocity gradients measured in optical lines cannot be used to infer launching radii. The NUV domain, which probes a faster and more collimated inner part of the jet, is likely more reliable to measure jet
rotation from Doppler gradients.

\end{itemize}

\section*{Acknowledgments}

\noindent{FL acknowledges support from the Joint Committee ESO government of Chile.  }

\noindent{FL thanks Sim\'on Casassus (Universidad de Chile) for his careful reading of the paper and numerous comments that led to extensive discussions among the authors. LB acknowledges support by CONICYT Grant PFB-06.}    

\noindent{This paper makes use of the following ALMA data: ADS/JAO.ALMA\#2012.1.00857.S. ALMA is a partnership of ESO (representing its member states), NSF (USA) and NINS (Japan), together with NRC (Canada), NSC and ASIAA (Taiwan), and KASI (Republic of Korea), in cooperation with the Republic of Chile. The Joint ALMA Observatory is operated by ESO, AUI/NRAO and NAOJ.}

\noindent{The National Radio Astronomy Observatory is a facility of the National Science Foundation operated under cooperative agreement by Associated Universities, Inc.}


\bibliographystyle{aa}
\bibliography{fab}

\begin{appendix} 
\section{Jet inclination }
\label{a:incli}

We revisit here previous estimates of the jet inclination to the line of sight. We
use our adopted distance of d=185\,pc and consistent line-of-sight and plane-of-sky velocities for the distant knots. We assumed these knots to be terminal bow shocks, in which the pattern of peaks luminosity move together with the matter within.

Along the PA of 98$^{\circ}$, a few HH knots were identified, especially one on the east side and one on the west side of the central source, designated as
HH 228 E1, and W \citep{krautter86,graham88}. Line-of-sight velocities ($V_{\rm rad}$) of -62 km.s$^{-1}$ and +38 km.s$^{-1}$ were reported for the E1 and W knots, respectively \citep{comeron10}. The \object{Th\,28} jet was observed at different epochs, which allowed \cite{wang09} to derive proper motions for the knots mentioned above. They found a proper motion of 0.47$\arcsec$yr$^{-1}$ for the E1 knot, and 0.37$\arcsec$yr$^{-1}$ for the W knot.

At the adopted distance for \object{Th\,28} of 185$\pm^{11}_{10}$\,pc, it corresponds to proper motions of 412$\pm$25\,km.s$^{-1}$ and 324$\pm$20\,km.s$^{-1}$ for the E1 and W knots, respectively. This provides independent measurements of the inclination of the system via inclination = $arctan\left(\dfrac{V_{\rm rad}}{\rm proper~motion}\right)$, of 81.3$\pm$1$^{\circ}$ and 83.2$\pm$1$^{\circ}$ for the E1 and W knots, respectively. We therefore adopt a global inclination of the jet, from the knots analysis, of 82.3$\pm$2$^{\circ}$.

\end{appendix}

\end{document}